\documentclass[useAMS,usenatbib,usegraphicx]{mn2e}
\title[Radio Emission from Cosmic Rays] {Radio Mini-Halo Emission from
Cosmic Rays in Galaxy Clusters and Heating of the Cool Cores}
\author[Y. Fujita and Y. Ohira]{Yutaka Fujita$^{1}$\thanks{E-mail:
fujita@vega.ess.sci.osaka-u.ac.jp} and Yutaka Ohira$^{2}$\\
$^{1}$Department of Earth and Space Science, Graduate School of Science,
Osaka University, 1-1 Machikaneyama-cho, \\ Toyonaka, Osaka 560-0043,
Japan\\ $^{2}$Department of Physics and Mathematics, Aoyama Gakuin
University, Fuchinobe, Chuou-ku, Sagamihara 252-5258, Japan}
\begin{document}

\date{Accepted 1988 December 15. Received 1988 December 14; in original form 1988 October 11}

\pagerange{000--000} \pubyear{0000}

\maketitle

\label{firstpage}

\begin{abstract}
 It has been proposed that the cool cores of galaxy clusters are stably
 heated by cosmic rays (CRs). If this is the case, radio mini-halos,
 which are often found in the central regions of cool core clusters, may
 be attributed to the synchrotron emission from the CRs. Based on this
 idea, we investigate the radial profiles of the mini-halos. First,
 using numerical simulations, we confirm that it is appropriate to
 assume that radiative cooling of the intracluster medium (ICM) is
 balanced with the heating by CR streaming. In these simulations, we
 assume that the streaming velocity of the CRs is the sound velocity of
 the ICM, and indicate that the heating is even more stable than the
 case where the streaming velocity is the Alfv\'en velocity. Then,
 actually assuming the balance between cooling and heating, we estimate
 the radial profiles of CR pressure in six clusters only from X-ray
 observations. Since the CR protons interact with the ICM protons, we
 can predict the radial profiles of the resultant synchrotron
 radiation. We compare the predictions with the observed radial profiles
 of the mini-halos in the six clusters and find that they are consistent
 if the momentum spectra of the CRs are steep. These results may
 indicate that the cores are actually being heated by the CRs. We also
 predict broad-band spectra of the six clusters, and show that the
 non-thermal fluxes from the clusters are small in hard X-ray and
 gamma-ray bands.
\end{abstract}


\section{Introduction}

Radio halos are diffuse radio emission that permeates clusters of
galaxies \citep{fer08a}. They are believed to be synchrotron radiation
from cosmic rays (CRs) in the intracluster medium (ICM). Although their
surface brightness is low, they are often extended on scales of
$\sim$Mpc. Since they are generally found in merging clusters, some
phenomena associated with cluster mergers, such as shocks and turbulence
in the ICM, may be responsible for the acceleration of the CR particles
\citep*[e.g.][]{jaf77a,roe99,tak00,fuj01,ohn02,fuj03,bru04,bru11}.

Among the radio-halos, so-called mini-halos are exceptional. While they
are small in number, they are found in the cool cores of cool core
clusters, which have not been disturbed by cluster mergers for a long
time \citep[e.g.][]{gov09a,mur09a}. Their size ($\sim 100$~kpc) is small
compered with ordinary radio halos. The relation between the radio
luminosity and the size is less clear for the mini-halos compared with
the ordinary halos \citep{mur09a}, which has made it difficult to find
the origin of the mini-halos. \citet{git02} proposed that the radio
synchrotron emission comes from CR electrons reaccelerated by turbulence
in the ICM.  \citet{pfr04} argue that the electrons illuminating the
mini-halos are of secondary origin; they are created through the
interaction between CR protons and ICM protons. \citet{fuj07c} also
proposed that the CR protons producing the secondary electrons are
accelerated at shocks around the central active galactic nuclei (AGNs)
in clusters.

For cool cores, there is another mystery with the ICM.  The radiative
cooling time of the ICM in the cores is generally less than the age of
the clusters, while that in the surrounding region is larger than the
age. If there are no heating sources, the ICM in the core should cool
and cannot sustain the pressure from the surrounding. In this case, a gas
flow toward the cluster centre should develop (a cooling flow; see
\citealt{fab94}). However, X-ray observations have shown that the
massive cooling flows are not developing in cluster cores
\citep[e.g.][]{ike97,mak01,pet01,tam01,kaa01,mat02}. Thus, some unknown
heating sources prevent cooling flows from developing in the cores. 

In cool cores, active AGNs interacting with the ICM are often observed
\citep*[e.g.][]{fab00,mcn00,bla01,mcn01,maz02,fuj02,joh02,kem02,tak03,fuj04}. Through
the AGN activities, CRs may be accelerated around the AGNs. The CRs may
transfer the energy generated by the AGNs to the surrounding ICM
\citep*[e.g.][]{tuc83,rep87,rep95,col04,pfr07,jub08}. One of the
transfer mechanisms is CR streaming \citep*{rep79,boh88,loe91,guo08}, in
which CRs interacting with Alfv\'en waves move outwards in the
cluster. The ICM is heated by the $PdV$ work of the CRs. We have shown
that the CRs can actually heat the ICM \citep[][hereafter
Paper~I]{fuj11}. The heating is fairly stable because it is not
localised around the AGN, and the density dependence of the heating term
is globally similar to the radiative cooling term. Moreover, the CR
distribution is not much affected by the temporal change of the ICM
distribution.

If CR protons are heating cores, they should produce non-thermal
emissions through their interaction with the thermal protons in the
ICM. In that case, the radio mini-halos could be closely related to the
heating of the cores. We calculated non-thermal spectra of cool cores
that are heated by the CR protons \citep[][hereafter
Paper~II]{fuj12}. We compared the results with the radio observations of
the mini-halo in the Perseus cluster. We found that the momentum
spectrum of the CRs must be steep so as to be consistent with the
observations of the mini-halo. We also showed that the detection of
non-thermal emissions from the CRs would be difficult in the hard X-ray
and gamma-ray bands, because of the steep spectrum. Thus, it is
desirable to find a way of studying the CR heating in more detail only
in the radio band.

In this paper, we show that CR heating can be studied closely by the
radial profiles of mini-halos. If the CRs are stably heating a cool
core, radiative cooling must be balanced with the heating at each
radius.  Assuming they balance, we can determine the radial profile of
the CR pressure using only X-ray data for a cluster. Once we obtain the
CR pressure profile, we can predict the radial profile of synchrotron
radiation from the CRs and we can compare the predicted profile with the
observed one. The idea of using the balance between the radiative
cooling and the CR heating has been proposed by \citet{col08a}. They
considered Coulomb and hadronic interactions as the CR heating. However,
since it has been indicated that CR streaming is a much more effective
heating source \citep[][Paper~I]{guo08}, we consider the CR streaming as
the heating source in cores.

This paper is organised as follows. In Section~\ref{sec:model}, we
describe our model on the estimation of CR pressure and the non-thermal
emissions from the CRs. In Section~\ref{sec:num}, we check whether the
CR heating is actually balanced with the radiative cooling using the
results of numerical simulations. In that section, we also show that the
CR heating is even more stable if the streaming velocity of the CRs is
the sound velocity of the ICM, compared to the case where the streaming
velocity is the Alfv\'en velocity. In Section~\ref{sec:comp}, we predict
radial profiles of mini-halos and compare them with the
observations. Finally, Section~\ref{sec:conc} is devoted to discussion
and conclusions. Throughout this paper we assume a $\Lambda$CDM
cosmology with $H_0=71\rm\: km\: s^{-1}\: Mpc^{-1}$, $\Omega_m=0.27$,
and $\Omega_\Lambda=0.73$. We consider protons as CRs unless otherwise
mentioned.

\section{Models}
\label{sec:model}

\subsection{CR Pressure}
\label{sec:Pc}

If the heating via CR streaming is balanced with radiative cooling in a
cool core, it must be
\begin{equation}
 \label{eq:balance}
H_{\rm st} \approx
n_e^2 \Lambda(T,Z)\:,
\end{equation}
where $n_e$ is the electron density of the ICM. The heating rate of the
CR streaming is given by
\begin{equation}
\label{eq:Hst}
 H_{\rm st} = -v_{\rm st}\frac{\partial P_c}{\partial r}\:,
\end{equation}
where $v_{\rm st}$ is the streaming velocity of the CRs, and $P_c$ is
the CR pressure (Paper~I). In our models, $\partial P_c/\partial
r<0$. We approximate the cooling function $\Lambda$ by
\begin{eqnarray}
 \Lambda(T,Z)&=&2.41\times 10^{-27}
\left[0.8+0.1\left(\frac{Z}{Z_\odot}\right)\right]
\left(\frac{T}{\rm K}\right)^{0.5}\nonumber\\
& &+ 1.39\times 10^{-16}
\left[0.02+0.1\left(\frac{Z}{Z_\odot}\right)^{0.8}\right]\nonumber\\
& &\times\left(\frac{T}{\rm K}\right)^{-1.0}\rm\: erg\: cm^3\:,
\end{eqnarray}
where $T$ is the temperature and $Z$ is the metal abundance of the ICM.
This function approximates the one derived by \citet{sut93a} for
$T\ga 10^5$~K and $Z\la 1\: Z_\odot$.

From equation~(\ref{eq:balance}), we obtain CR pressure:
\begin{equation}
\label{eq:Pc}
 P_c(r)=\int_r^{r_{\rm max}}\frac{n_e^2 \Lambda}{v_{\rm st}}dr\:,
\end{equation}
where $r_{\rm max}$ is the radius where $P_c$ approaches zero. Since
$n_e^2$ rapidly decreases outside the core, the integration is
insensitive to the value if we assume that it is much larger than the
core radius. Thus, we assume that $r_{\rm max}=1$~Mpc.

CRs move outwards as a whole in the cluster with waves that scatter the
CRs. If the waves are Alfv\'en waves, the streaming velocity $v_{\rm
st}$ may be the Alfv\'en velocity $v_A$. However, it has been indicated
that the CR streaming velocity may be much larger than the Alfv\'en
velocity in hot ICM, because in hot plasma the Alfv\'en waves may suffer
strong resonant damping at small wave lengths by thermal protons. In
this case, the sound velocity $c_s$ may be appropriate as the streaming
velocity \citep*{hol79,ens11}. In this paper, we mainly consider the
case of $v_{\rm st}=c_s$.

In this study, we do not discuss the injection of CRs into the ICM in
detail. They may be directly injected at shocks \citep{fuj07c}, or they
may be transported by buoyant bubbles \citep{guo08}. In the latter case,
the bubbles are created through AGN activities, which also accelerate
CRs around the bubbles. The bubbles contain the CRs and carry them from
the AGN out to large distances. As the bubbles adiabatically rise, the
CRs may escape from the bubbles into the ICM or they may be injected
into the ICM through the shredding of the bubbles by Rayleigh-Taylor and
Kelvin-Helmholtz instabilities. While the adiabatic expansion reduces
the energies of CRs, it does not change the slope of the momentum
spectrum. Observations have shown that the bubbles with a size of $\sim
10$~kpc are often found at the centre of cores, while large scale ($\sim
100$~kpc) shocks appear to be rare \citep[e.g.][]{bir04a,mcn12a}. This
may suggest that CR injection via bubbles is more common.

\subsection{Synchrotron Emission from CRs}
\label{sec:nonth}

Although we can obtain $P_c(r)$ from equation~(\ref{eq:Pc}) for given
temperature and density distributions and the streaming velocity, we
need to specify the momentum spectrum of the CRs to calculate the
synchrotron spectrum of the CRs. For that purpose, we adopt the
following spectrum:
\begin{equation}
\label{eq:Npr}
 N(p,r)= A_{\rm cr}(r) p^{-x}e^{-p/p_{\rm max}}\:,
\end{equation}
where $p$ is the CR momentum, $x$ is the index, and $p_{\rm max}$ is the
cutoff momentum. 

The hadronic loss of CR protons could modify the slope of the spectrum
\citep[e.g.][]{bru04,ens07a,guo08}. In our case, the escape time of CRs
from the core ($\sim 10^8$--$10^9$~yr) is smaller than the cooling time
via hadronic loss. Thus, we ignore the modification of the spectrum by
the loss.

We assume that $x$ and $p_{\rm max}$ do not depend on $r$ (Paper~II). On
the other hand, the normalisation of the spectrum, $A_{\rm cr}(r)$,
depends on $r$ and it is determined from the following equation for
given $P_c(r)$:
\begin{equation}
\label{eq:Pc2}
 P_c(r) = \frac{c}{3}\int_{p_{\rm min}}^\infty
\frac{p^2 N(p,r)}{\sqrt{p^2 + m^2 c^2}} dp\:.
\end{equation}
We determine the cutoff momentum $p_{\rm max}$ assuming that the
high-energy CRs are accelerated at shocks formed through the activities
of the central AGN (see Paper~II). The cutoff momentum is determined by
the age of the shock, and $p_{\rm max}c\sim 10^5$~TeV if the energy
input through an explosive activity of the AGN is $\sim
10^{60}$~erg. Although the actual acceleration mechanism is not clear
(e.g. CRs may be accelerated nearby the black hole and form jets;
 \citealt{sik05a}), the following results are not sensitive to $p_{\rm
max}$. The minimum momentum, $p_{\rm min}$, will be set in
Section~\ref{sec:HC}.  On the other hand, as will be shown later, the
synchrotron spectrum and luminosity are sensitive to the index $x$.

Once we fix the momentum spectrum of the CRs at each radius, we can
calculate synchrotron and other non-thermal emissions from secondary
electrons that are generated through the interaction between CR protons
and ICM protons, and through the decay of charged pions. We can also
calculate $\pi^0$-decay gamma-rays. We use the radiation model of
\citet{fan07} and the proton-proton collision model of \citet{kar08};
the details are described in \citet{fuj09c}.

\section{Comparison with Numerical Simulations}
\label{sec:num}

\subsection{Balance between Heating and Cooling}
\label{sec:HC}

Before we study real clusters, we investigate whether the results of
numerical simulations are well approximated by
equation~(\ref{eq:balance}). We calculate $P_c(r)$ based on the models
in Papers~I and~II, although we make a few modifications as follows.  We
adopt the gravitational potential and the initial ICM profile of
Model~LCRs in Paper~II. This model is originally constructed to
reproduce the observations of the Perseus cluster. We do not include
thermal conduction, and the CR diffusion can be ignored. The CR streaming
velocity $v_{\rm st}$ is the sound velocity $c_s$. We reevaluate the
parameters for Coulomb and hadronic losses of CR protons. This is
because although we found that the index must be large ($x\sim 3$) in
Paper~II, the model we used to obtain $P_c(r)$ in Paper~II did not
include the effect. The parameters of the losses are calculated as
follows \citep[see also][]{guo08}.

The energy loss of a proton with velocity of $v=\beta c$ by Coulomb loss
is given by \citet{man94a}:
\begin{equation}
\label{eq:Cou}
 \left(\frac{d E_p}{dt}\right)_c = -4.96\times 10^{-19}
\left(\frac{n_c}{\rm cm^{-3}}\right)\frac{\beta^2}{\beta^3 + x_m^3}
\rm\: erg\: s^{-1}\:,
\end{equation}
where $x_m=0.0286\:(T/2\times 10^6\rm\: K)^{1/2}$. For the momentum
spectrum~(\ref{eq:Npr}), the overall Coulomb loss rate is
\begin{eqnarray}
\label{eq:gamc}
 \Gamma_c&=&\int_{p_{\rm min}}^{\infty}
N\left(\frac{d E_p}{dt}\right)_c dp\nonumber\\
&=&-7.3\times 10^{-16}\left(\frac{n_e}{\rm cm^{-3}}\right)\nonumber\\
& &\times\left(\frac{e_c}{\rm erg\: cm^{-3}}\right)
\rm\: erg\: s^{-1}\: cm^{-3} \:,
\end{eqnarray}
where $e_c$ is the CR energy density. In this subsection, we assume that
$x=3$ and $T=8$~keV, which are the typical values for the clusters we
later investigate. The minimum momentum is set to be $p_{\rm
min}c=137$~MeV, which corresponds to the kinetic energy of 10~MeV and
was adopted by \citet{guo08}. Around this momentum, the Coulomb loss of
a proton is maximum ($\beta\sim x_m$ in equation~[\ref{eq:Cou}]).

The energy-loss rate of a CR proton due to pion production is
approximately given by
\begin{equation}
 \left(\frac{d E_p}{dt}\right)_h \approx -0.5
n_N \sigma_{pp} c T_p \theta (p - p_{\rm thr})\:,
\end{equation}
where $n_N$ is the nucleon density, $\sigma_{pp}$ is the $pp$
cross-section, $p_{\rm thr}$ is the threshold momentum, $T_p$ is the
kinetic energy of the proton, and $\theta$ is the step-function
\citep{ens07a}. For $n_N$, $\sigma_{pp}$, and $p_{\rm thr}$, we adopted
the ones used in \citet{ens07a}. The overall hadronic loss rate is
\begin{eqnarray}
\label{eq:gamh}
 \Gamma_h&=&\int_{p_{\rm thr}}^{\infty}
N\left(\frac{d E_p}{dt}\right)_p dp\nonumber\\
&=&-1.5\times 10^{-17}\left(\frac{n_e}{\rm cm^{-3}}\right)\nonumber\\
&&\times\left(\frac{e_c}{\rm erg\: cm^{-3}}\right)
\rm\: erg\: s^{-1}\: cm^{-3} \:.
\end{eqnarray}
Thus, the energy loss rate is $\Gamma_{\rm loss}=-\Gamma_c - \Gamma_h =
\zeta_c n_c e_c$, where $\zeta_c=7.5\times 10^{-16}\rm\: cm^3\:
s^{-1}$. Since $\sim 1/6$ of the inelastic energy goes into secondary
electrons during hadronic collisions \citep{man94a,guo08}, the heating
rate by the Coulomb and hadronic collisions is $H_{\rm coll}=-\Gamma_c -
\Gamma_h/6=\eta_c n_c e_c$, where $\eta_c=7.4\times 10^{-16}\rm\: cm^3\:
s^{-1}$. We use these values in equations~(3) and (4) in Paper~II. The
values are not much different from the ones derived by \citet{guo08} and
adopted by us for $x=2.4$ using a slightly different formula
($\zeta_c=7.51\times 10^{-16}\rm\: cm^3\: s^{-1}$ and $\eta_c=2.63\times
10^{-16}\rm\: cm^3\: s^{-1}$).

\begin{figure}
\includegraphics[width=84mm]{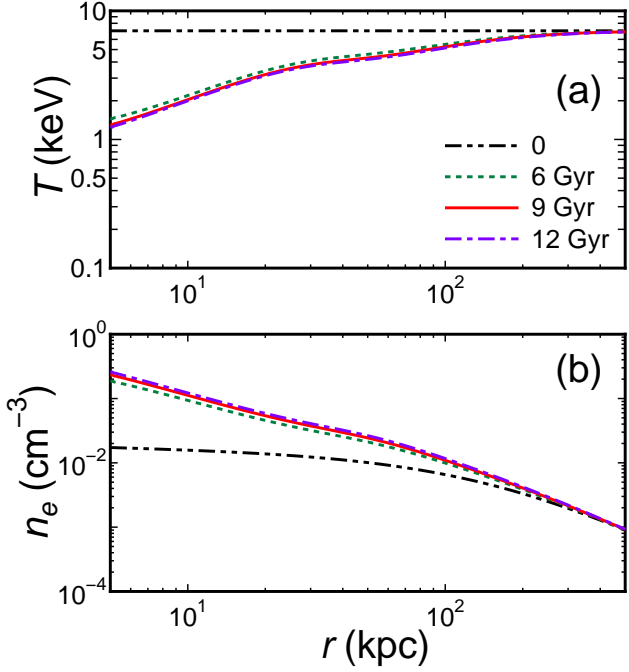} \caption{(a) Temperature and (b)
density profiles for Model~1. \label{fig:Tn_lcrg} }
\end{figure}

\begin{figure}
\includegraphics[width=84mm]{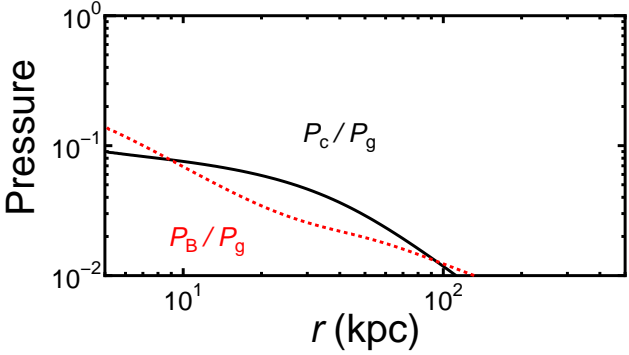} 
\caption{Profiles of the ratios
$P_c/P_g$ (solid) and $P_B/P_g$ (dotted) at $t=9$~Gyr for Model~1.}
\label{fig:Pcb_lcrg}
\end{figure}

\begin{figure}
\includegraphics[width=74mm]{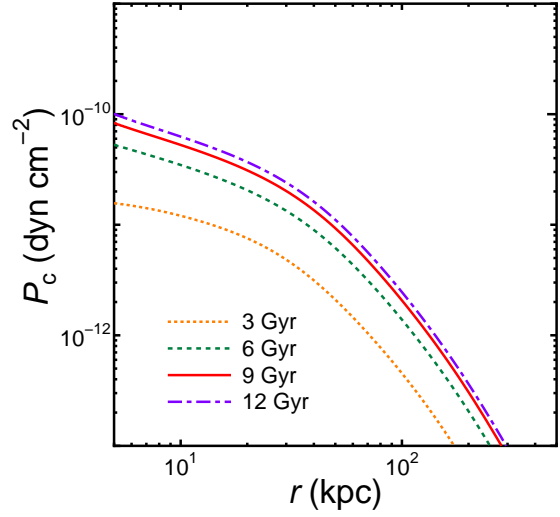} 
\caption{CR pressure profiles for Model~1.
 \label{fig:Pc_lcrg}}
\end{figure}

\begin{figure}
\includegraphics[width=84mm]{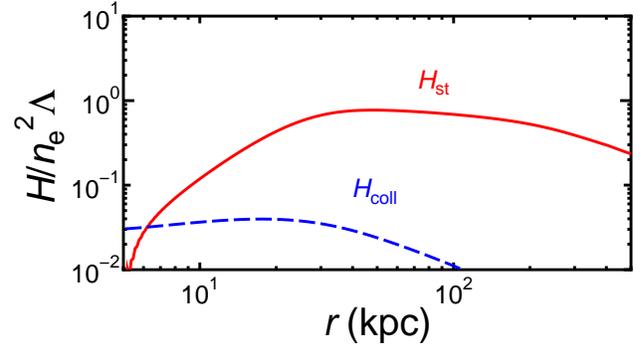} 
\caption{Relative importance of CR streaming ($H_{\rm st}$) and
collisional heating ($H_{\rm coll}$) at $t=9$~Gyr for Model~1.
\label{fig:heat_lcrg}}
\end{figure}

\begin{figure}
\includegraphics[width=84mm]{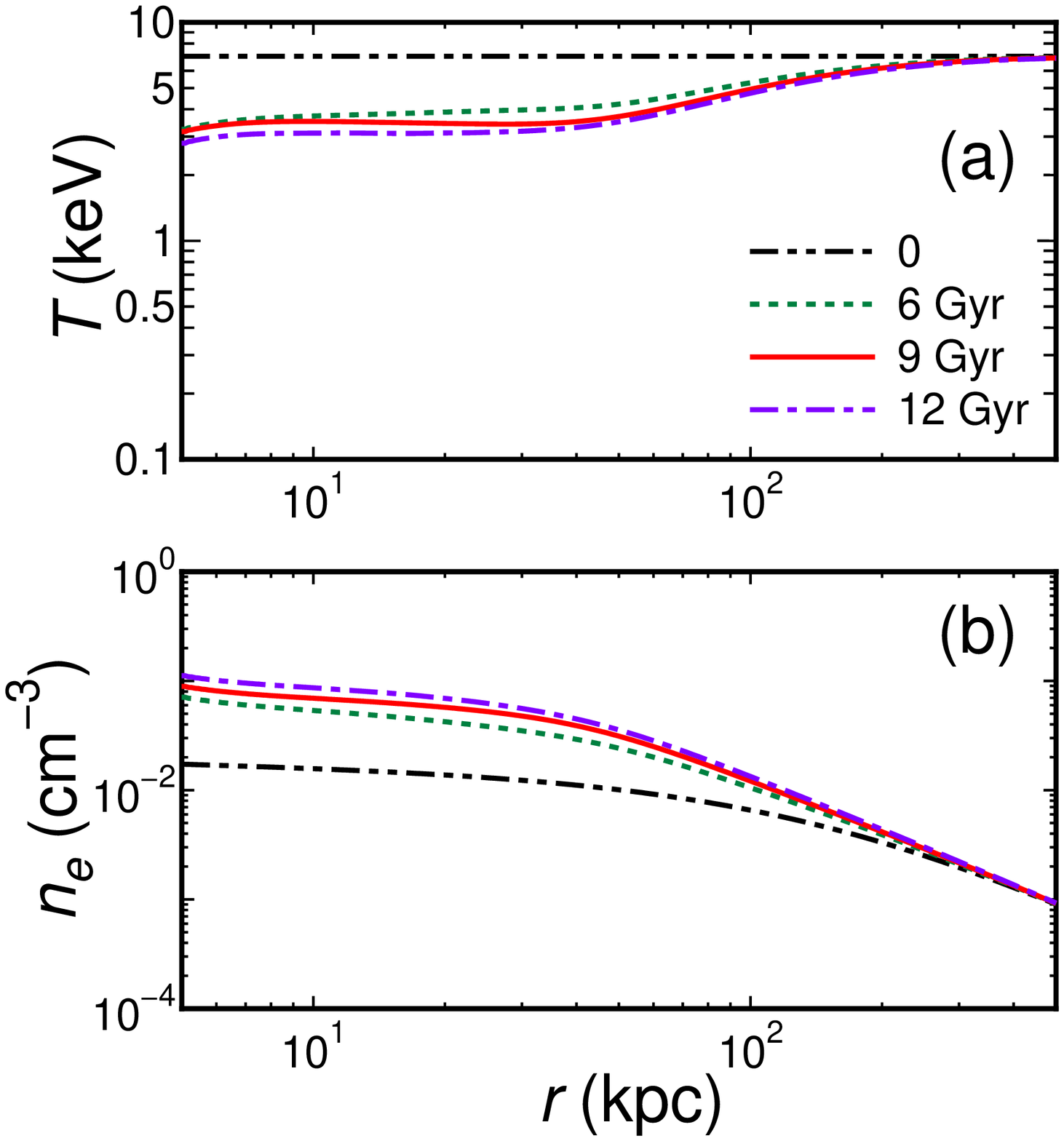}
\caption{Same as Figure~\ref{fig:Tn_lcrg} but for Model~2.
\label{fig:Tn_lcrh}}
\end{figure}

\begin{figure}
\includegraphics[width=84mm]{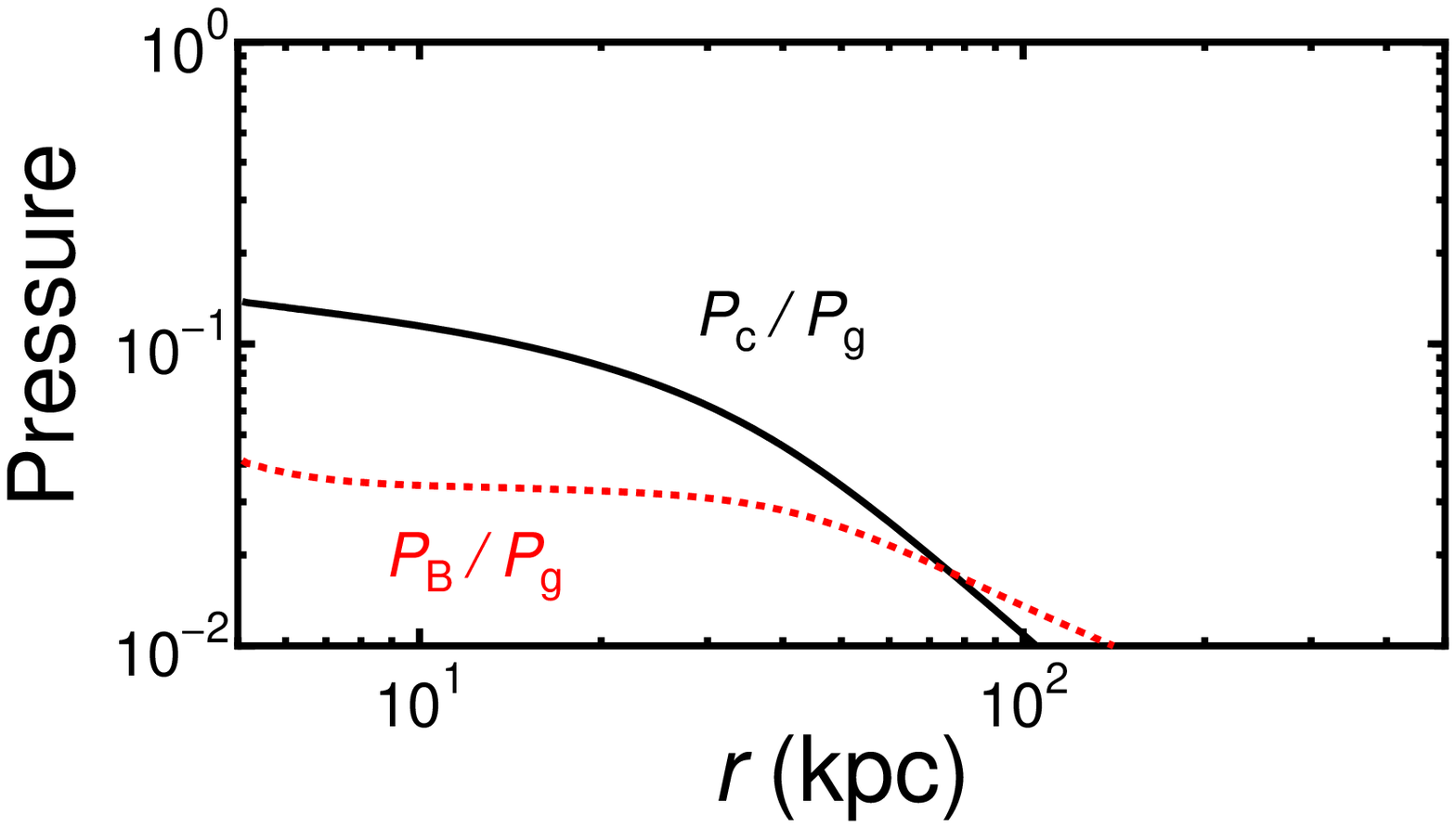} 
\caption{Same as Figure~\ref{fig:Pcb_lcrg} but for Model~2.
\label{fig:Pcb_lcrh}}
\end{figure}

\begin{figure}
\includegraphics[width=74mm]{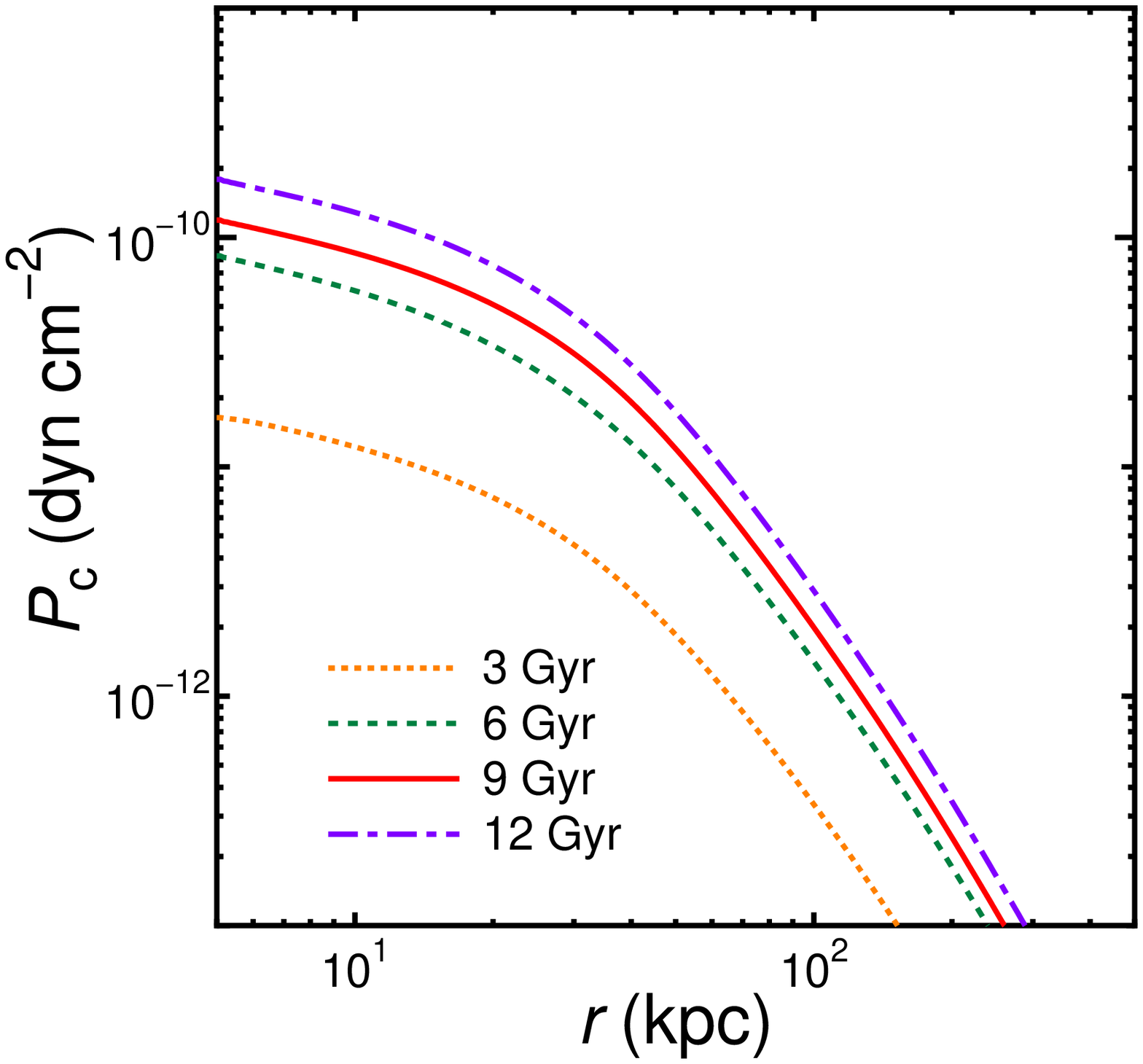} 
\caption{Same as Figure~\ref{fig:Pc_lcrg} but for Model~2.
\label{fig:Pc_lcrh}}
\end{figure}

\begin{figure}
\includegraphics[width=84mm]{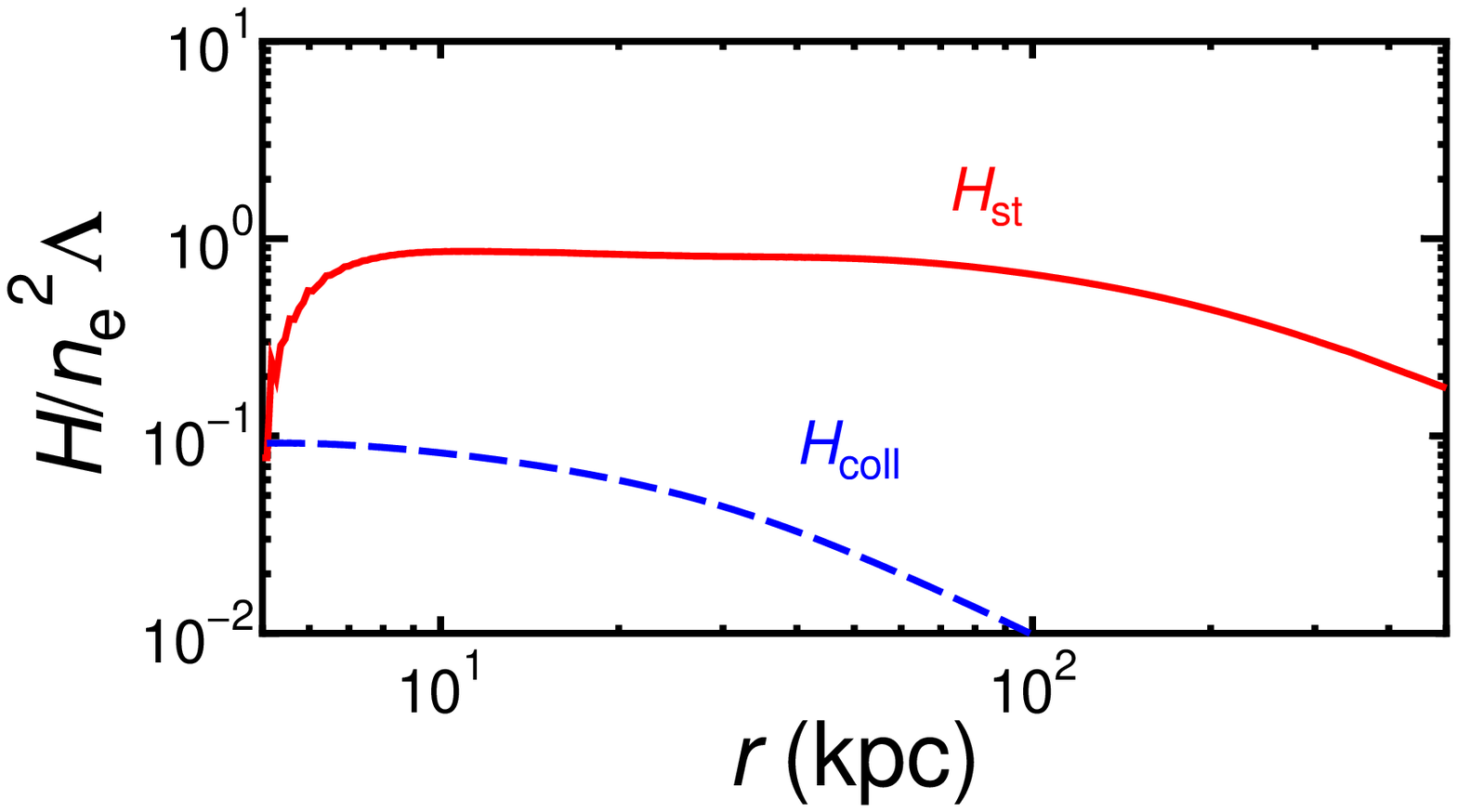} 
\caption{Same as Figure~\ref{fig:heat_lcrg} but for Model~2.
\label{fig:heat_lcrh}}
\end{figure}

\begin{figure}
\includegraphics[width=84mm]{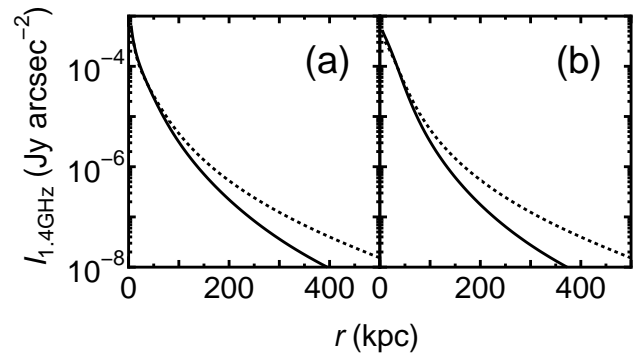} \caption{Profiles of synchrotron
radiation when we use $P_c(r)$ directly calculated via numerical
simulations (solid lines) and when we use $P_c(r)$ estimated by
equation~(\ref{eq:Pc}) (dotted lines). (a) Model~1 and (b) Model~2. The
redshift of the cluster is set to be $z=0.0179$ (the Perseus cluster).
\label{fig:syn}}
\end{figure}

\begin{figure}
\includegraphics[width=84mm]{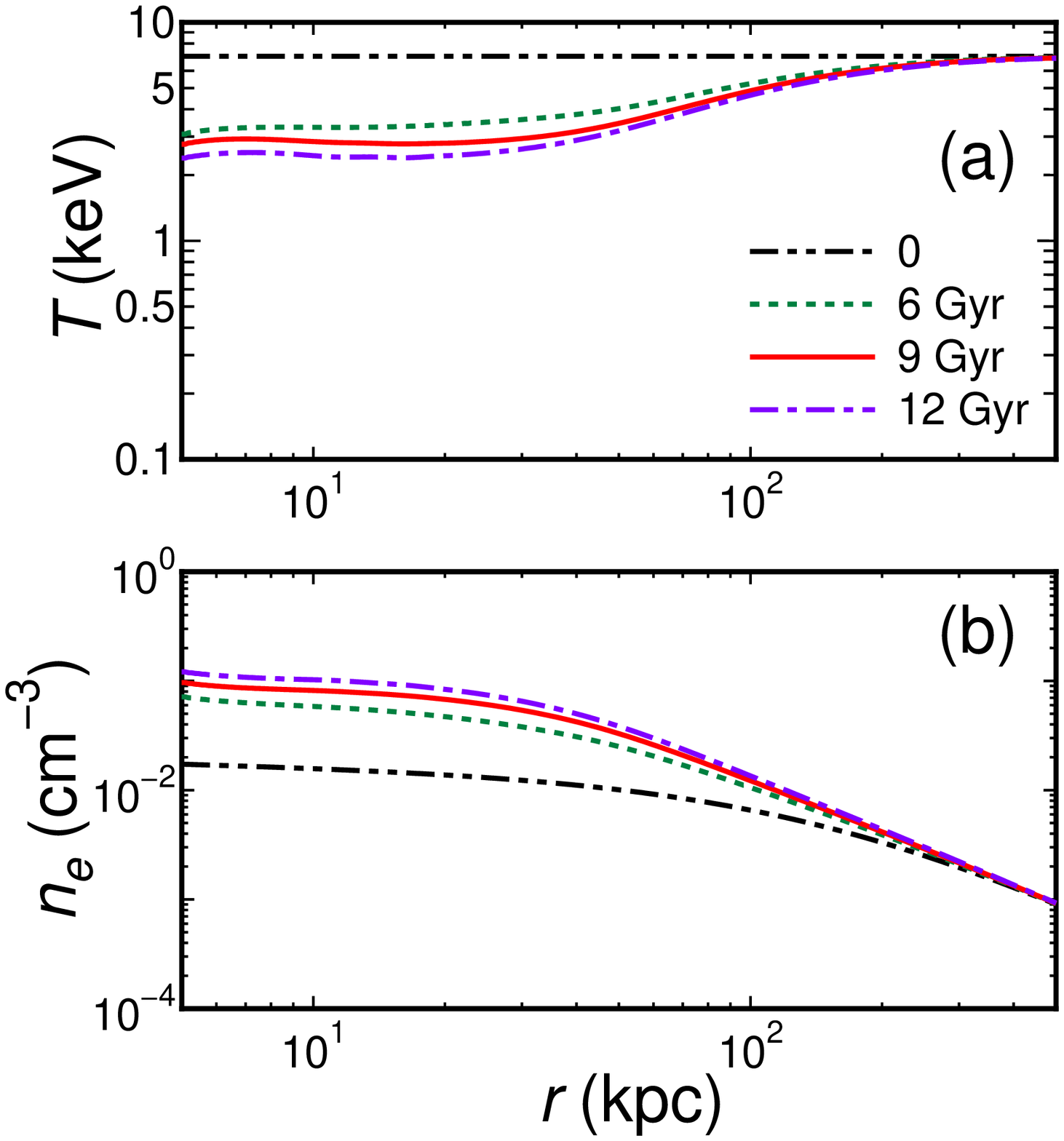}
\caption{Same as Figure~\ref{fig:Tn_lcrg} but for Model~3.
\label{fig:Tn_lcrh3}}
\end{figure}

\begin{figure}
\includegraphics[width=84mm]{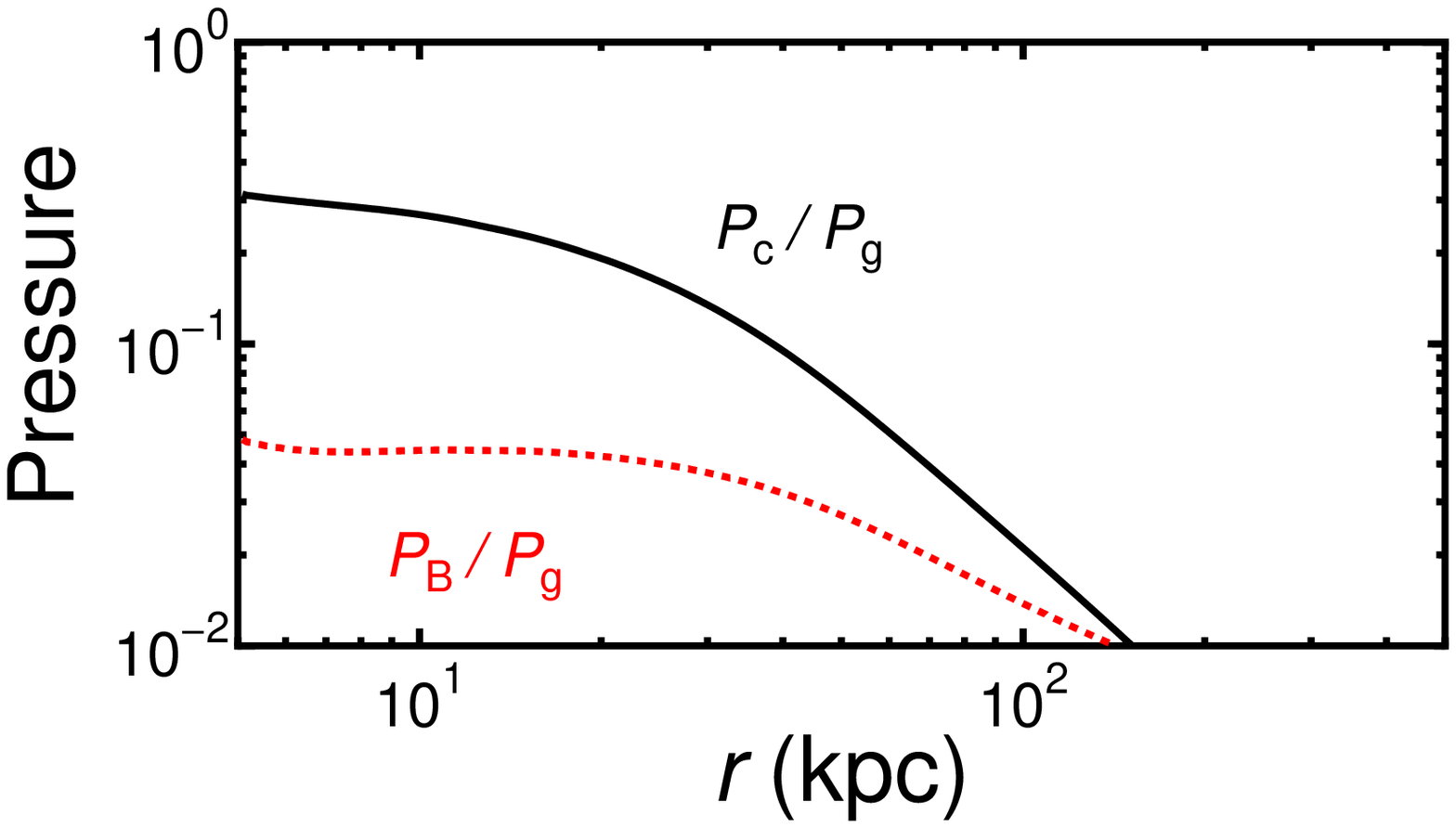} 
\caption{Same as Figure~\ref{fig:Pcb_lcrg} but for Model~3.
\label{fig:Pcb_lcrh3}}
\end{figure}

\begin{figure}
\includegraphics[width=74mm]{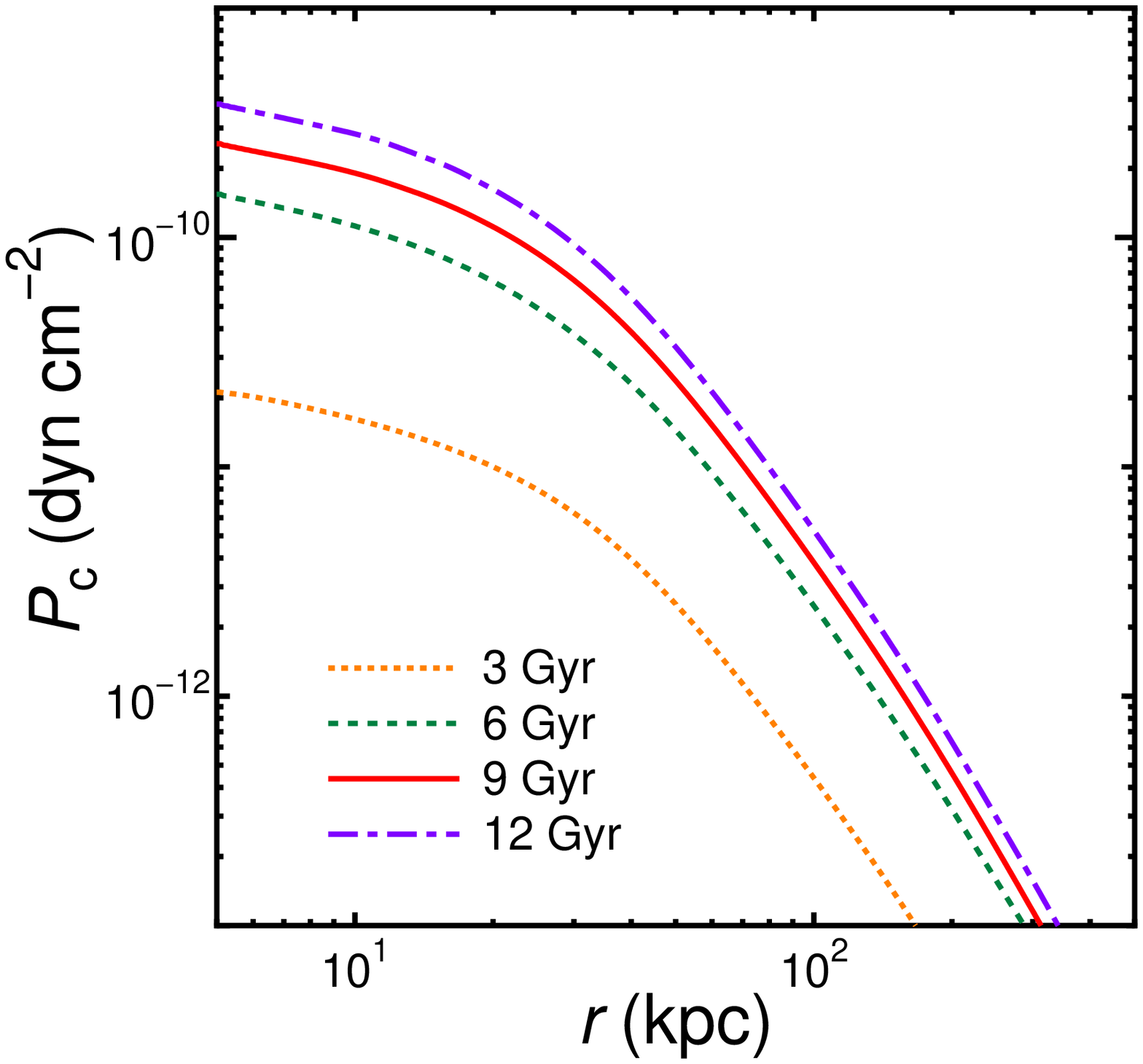} 
\caption{Same as Figure~\ref{fig:Pc_lcrg} but for Model~3.
\label{fig:Pc_lcrh3}}
\end{figure}

\begin{figure}
\includegraphics[width=84mm]{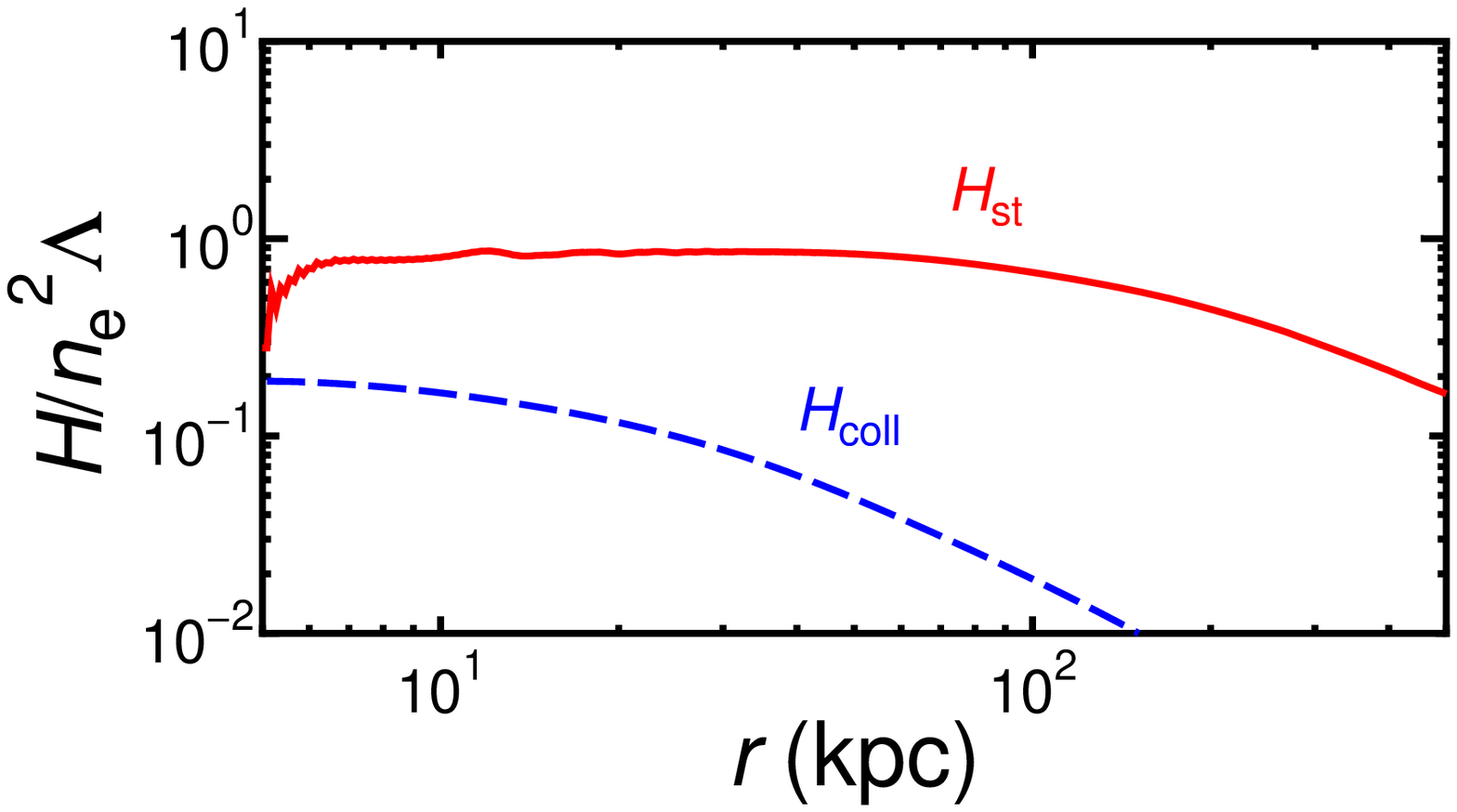} 
\caption{Same as Figure~\ref{fig:heat_lcrg} but for Model~3.
\label{fig:heat_lcrh3}}
\end{figure}

In this subsection, we consider two models, which are based on
Model~LCRs in Paper~II. Contrary to Model~LCRs, we adopted the above
$\zeta_c$ and $\eta_c$. Moreover, we assume that the adiabatic index of
the CRs is $\gamma_c=5/3$ instead of 4/3 as an extreme case, because the
index approaches 5/3 if most of the CRs have low energies. In Papers~I
and~II, the total CR injection rate by the AGN was given by $L_{\rm
AGN}=\epsilon \dot{M} c^2$, where $\dot{M}$ is the inflow rate of the
gas toward the cluster centre. If the CRs are gradually injected into
the ICM as the buoyant bubbles rise, their injection should not be
localised. Thus, we give the CR injection rate per unit volume by
$\dot{S}_c\propto r^{-\nu}$ for $20\la r \la 150$~kpc. The lower bound
is based on the observations of the bubbles \citep{bir04a,guo08}.
Although the upper bound is uncertain, the results are not sensitive to
the value, because $r^3\dot{S}_c$ is a decreasing function of $r$ if
$\nu>3$. In one model, we assume that the efficiency is
$\epsilon=2.5\times 10^{-4}$, and the radial dependence is
$\nu=3.1$. These values are the same as those in Model~LCRs in
Paper~II. We refer to this model as Model~1. We consider another model,
which is the same as Model~1 but the AGN feedback is stronger
(Model~2). We assume that $\epsilon=2.5\times 10^{-3}$, and that the
energy input is more centrally concentrated ($\nu=3.5$). We have shown
that the CR heating is locally unstable with these parameters
($\epsilon$ and $\nu$) if the streaming velocity is the Alfv\'en
velocity and if thermal conduction is not included, because the heating
is too strong and too centrally concentrated (Paper~I). We assume that
$B=10\: (n_e/0.016\rm\: cm^{-3})^{2/3}\rm\: \mu G$, which is the same as
that in Model~LCRs in Paper~II.

It is to be noted that $\dot{M}$ represents the mass flow that passes
the inner boundary ($r=5$~kpc), and that not all the mass is swallowed
by the central black hole. Most of the gas may be consumed by star
formation in the central galaxy \citep[e.g.][]{mcn12a}. For example, if
2.5\% of $\dot{M}$ goes to the neighbourhood of the black hole, and 10\%
of the rest mass energy is converted to CRs, the efficiency is
$\epsilon=2.5\times 10^{-3}$. Although the conversion rate to CRs (10\%)
may be rather high, recent studies have shown that most of the AGNs at
present seem to be in radiatively inefficient mode or radio mode
\citep[e.g.][]{mcn12a}, and that the inertia of AGN jets seems to be
dominated by CR protons \citep{sik05a}. Thus, the conversion rate may
not be unrealistic. The injected CRs eventually stream out of the
cluster, to beyond the virial radius.

Figures~\ref{fig:Tn_lcrg}--\ref{fig:heat_lcrg} are the results for
Model~1. Figures~\ref{fig:Tn_lcrg} and~\ref{fig:Pcb_lcrg} are almost
identical to the corresponding figures for Model~LCRs in Paper~II
(Figures~8 and~9 in that paper). The cluster is isothermal and in
hydro-statistic equilibrium at $t=0$ (Figure~\ref{fig:Tn_lcrg}). The
mass inflow rate gradually increases with time. At $t=12$~Gyr, the mass
inflow rate is $\dot{M}=144\rm\: M_\odot\: yr^{-1}$. If there is no
heating, it should be $760\rm\: M_\odot\: yr^{-1}$ (Paper~I). In
Figure~\ref{fig:Pcb_lcrg}, we show the ratios of the CR pressure $P_c$
and the magnetic pressure $P_B$ to the ICM pressure $P_g$.
Figure~\ref{fig:Pc_lcrg} shows the evolution of $P_c(r)$. For $r\ga
100$~kpc, the density of CRs nearly follows $\rho_c\propto
P_c^{1/\gamma_c} \propto r^{-2}$, or $P_c\propto r^{-10/3}$. This is
because of the continuity, almost constant sound velocity (streaming
velocity), and inefficient cooling in the outer region of the
cluster. However, the slope is less steep for $r\la 100$~kpc, because
CRs are injected not at the cluster centre but mainly at $20\la r \la
150$~kpc.  Figure~\ref{fig:heat_lcrg} demonstrates that the CR heating,
$H_{\rm st}$, and radiative cooling, $n_e^2\Lambda$, are well balanced
except for the innermost and the outermost regions. The figure shows
that $0.5\la H_{\rm st}/(n_e^2\Lambda)<0.8$ for $20\la r\la
200$~kpc. Figures~\ref{fig:Tn_lcrh}--\ref{fig:heat_lcrh} demonstrate the
results for Model~2. The CR heating is stable in spite of the large
$\epsilon$ and $\nu$. This means that the CR heating with $v_{\rm
st}=c_s$ is even more stable than that with $v_{\rm st}=v_A$. Because of
the strong feedback, radiative cooling is almost cancelled out. At
$t=12$~Gyr, the mass inflow rate is only $\dot{M}=17\rm\: M_\odot\:
yr^{-1}$. The reason of the additional stability is that if the ICM
temperature decreases at the cluster centre, the streaming velocity
$v_{\rm st}=c_s$ decreases there. Thus, the escape of the CRs from the
centre delays, which increases $|dP_c/dr|$ and the heating rate
(equation~[\ref{eq:Hst}]). Figure~\ref{fig:heat_lcrh} shows that $0.7\la
H_{\rm st}/(n_e^2\Lambda)<1$ for $r\la 100$~kpc, and $H_{\rm
st}/(n_e^2\Lambda)\sim 0.5$ at $r\sim 200$~kpc. The well-established
balance between heating and cooling both for Models~1 and~2 suggests
that the ICM evolves in such a way as to achieve the balance regardless
of the details of the AGN heating model. In other words,
equation~(\ref{eq:balance}) is the most important factor that determines
the ICM profile in the core. It is to be noted that the bulk of CR
energy is dissipated as heat within the core. In Model~2 for example,
the CR energy injection rate from the AGN is $2\times 10^{45}\:\rm erg\:
s^{-1}$ at $t=12$~Gyr. At the same time, the flux of CR enthalpy
escaping the core (say, at $r=300$~kpc) is only $4\pi \gamma_c v_{\rm
st} r^2 P_c/(\gamma_c-1)\sim 3\times 10^{44}\:\rm erg\: s^{-1}$.

Figure~\ref{fig:syn} shows the radial profiles of synchrotron radio
emission for Models~1 and~2 at $t=9$~Gyr. We chose $x=2.9$, because the
value is favourable when compared with observations (see
Section~\ref{sec:rad}). We compare the profiles that are calculated
using $P_c(r)$ in Figures~\ref{fig:Pc_lcrg} and~\ref{fig:Pc_lcrh} with
those using $P_c(r)$ derived from $n_e(r)$, $T(r)$, and $v_{\rm st}(r)
(=c_s(r))$ by equation~(\ref{eq:Pc}). Figure~\ref{fig:syn} shows that
their differences are small in the core region ($r\la 200$~kpc), which
means that the approximation by equation~(\ref{eq:balance}) is
good. However, outside the core region, the results using
equation~(\ref{eq:Pc}) is overestimated by a factor of a few, because
the CR heating is not balanced with radiative cooling
(Figures~\ref{fig:heat_lcrg} and~\ref{fig:heat_lcrh}).

\begin{table*}
 \centering
 \begin{minipage}{140mm}
  \caption{Cluster profiles}
  \begin{tabular}{@{}lllc@{}}
  \hline
   Name  & $z$ & Profiles\footnote{The units for $n_e$, $T$, $Z$, 
and $r$ are
   $\rm cm^{-3}$, keV, $Z_\odot$, and kpc, respectively}  & Refs \\
 \hline
 Perseus & 0.0179 & 
$n_e(r)=\frac{0.0193}{1+(r/18)^3}
+ \frac{0.046}{[1+(r/56)^2]^{1.8}}
+ \frac{0.0048}{[1+(r/197)^2]^{0.87}}$ 
& \footnote{\citet{mat06}} \\
         &        &
$T(r) = 7.0
\frac{1+(r/70)^3}{2.3+(r/70)^3}$
& $b$ \\
         &        &
$Z(r) = 0.46 
+ 0.27\exp\left[-\frac{1}{2}\left(\frac{r}{51}\right)^2\right]
- 0.42\exp\left[-\frac{1}{2}\left(\frac{r}{11}\right)^2\right]$
& \footnote{\citet{chu03a}} \\
 A1835   & 0.2532 & 
$n_e(r)=\frac{0.062}{[1+(r/41)]^{0.74}}$ (for $r\leq 231$)
& \footnote{\citet*{maj02a}} \\
         &        &
$n_e(r)=\frac{0.017}{[1+(r/158)]^{1.06}}$ (for $r> 231$)
&     \\
         &        &
$T(r) = 7.8 
- 3.7\exp\left[-\frac{1}{2}\left(\frac{r}{48}\right)^2\right]$
& $d$ \\
         &        &
$Z(r) = 0.28 
+ 0.15\exp\left[-\frac{1}{2}\left(\frac{r}{57}\right)^2\right]$
& $d$ \\
 A2029   & 0.0765 & 
$n_e(r)^2
=(0.0168)^2\frac{(r/85.5)^{-1.164}}
{[1+(r/85.5)^2]^{1.05}} \frac{1}{[1+(r/923)^3]^{0.556}}
+ \frac{(0.376)^2}{[1+(r/5.08)^2]^3}$ 
& \footnote{\citet{vik06a}} \\
         &        &
$T(r)=16.19\frac{[(r/94.5)^{0.48}+0.10]}
{(r/94.5)^{0.48}+1}
\frac{(r/3088)^{0.03}}{[1+(r/3088)^{1.57}]^{3.76}}$
& $e$ \\
         &        &
$Z(r) = 0.27
+ 0.34\exp\left[-\frac{1}{2}\left(\frac{r}{96}\right)^2\right]$
& \footnote{\citet{vik05a}} \\
 A2390   & 0.2280 & 
$n_e(r)=0.00385\frac{(r/314)^{-0.9455}}
{[1+(r/314)^2]^{0.5142}}
\frac{1}{[1+(r/1223)^3]^{0.09383}}$ 
& $e$ \\
        &        &
$T(r)=19.34\frac{[(r/218)^{0.08}+0.12]}
{(r/218)^{0.08}+1}
\frac{(r/2507)^{0.10}}
{[1+(r/2507)^{5.0}]^{2.0}}$
& $e$ \\
         &        &
$Z(r) = 0.33
+ 0.42\exp\left[-\frac{1}{2}\left(\frac{r}{31}\right)^2\right]$
& $f$ \\
 RXJ1347   & 0.4510 & 
$n_e(r)=\frac{0.25}{[1+(r/23.5)]^{0.78}}$ 
& \footnote{\citet*{all02a}} \\
        &        &
$T(r)$ is the curve shown in Figure~2 of reference $h$
& \footnote{\citet*{git07b}} \\
         &        &
$Z(r) = 0.33$
& \footnote{\citet{ota08a}}  \\
 Ophiuchus   & 0.028 & 
$n_e(r)=\frac{0.0127}{[1+(r/23.1)]^{1.11}}
 + \frac{0.0101}{[1+(r/162)]^{1.11}}$ 
& \footnote{\citet{nev09a}} \\
        &        &
$T(r) = 9.3
- 3.0\exp\left[-\frac{1}{2}\left(\frac{r}{19}\right)^2\right]$
& $^{j,}$\footnote{\citet{fuj08d}} \\
         &        &
$Z(r) = 0.27 
+ 0.30\exp\left[-\frac{1}{2}\left(\frac{r}{27}\right)^2\right]$
& $j$,$k$  \\
\hline
\end{tabular}
\end{minipage}
\end{table*}

\begin{figure}
\includegraphics[width=84mm]{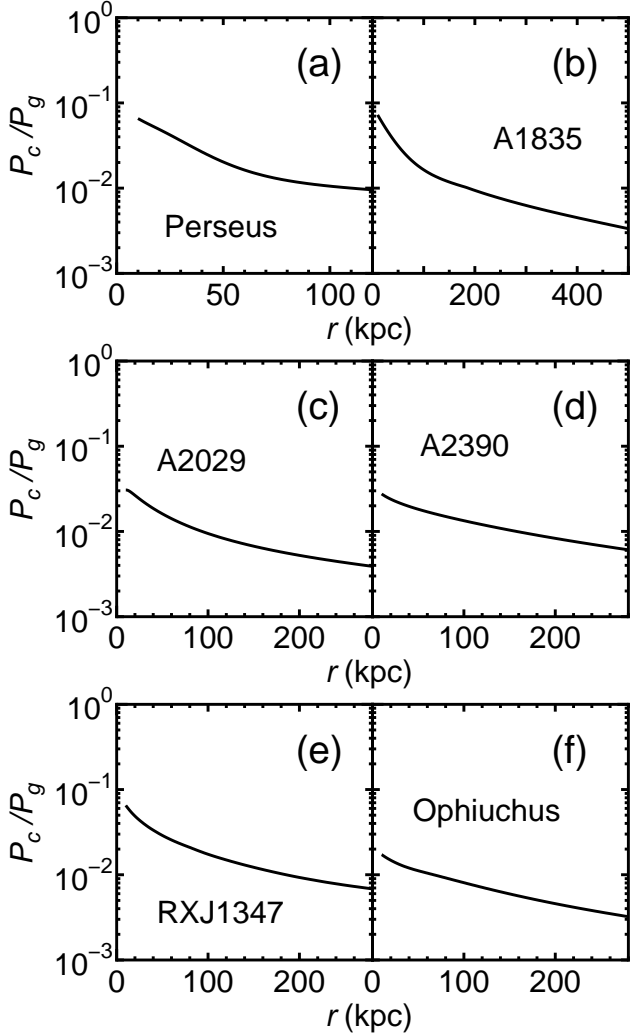} \caption{Ratio of CR pressure
 $P_c$ to ICM pressure $P_g$.  \label{fig:pp}}
\end{figure}

\begin{figure}
\includegraphics[width=84mm]{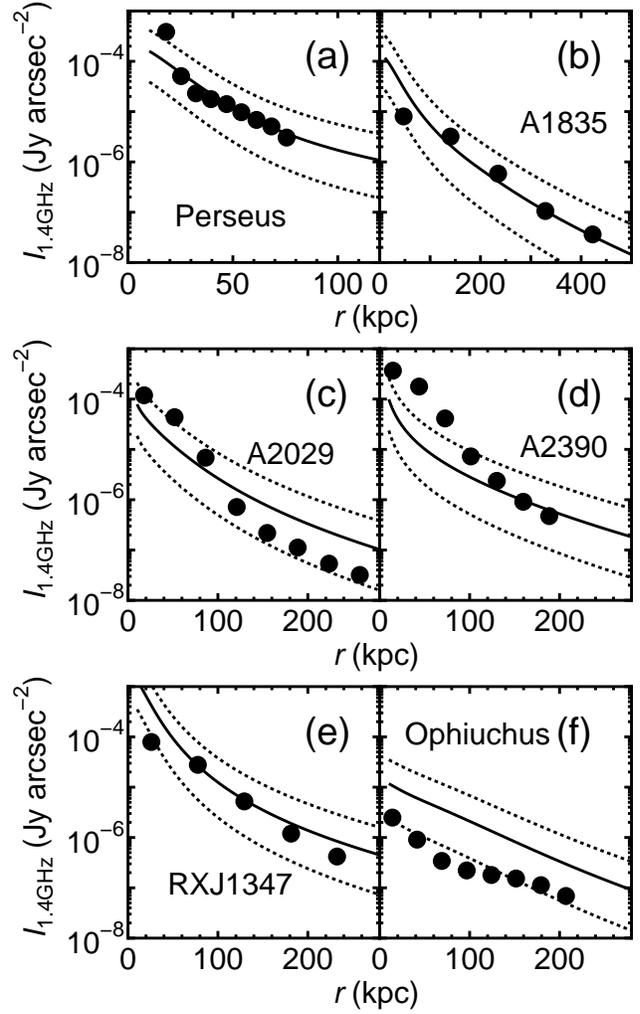} \caption{Predicted surface
brightness profiles of synchrotron emission of the six clusters at
$\nu=1.4$~GHz when $v_{\rm st}=c_s$ for $x=2.9$ (solid lines), 2.6
(upper dotted lines), and 3.3 (lower solid lines). Filled circles
are the observations \citep{ped90a,bac03a,git07c,mur09a}. Redshifts have
been corrected. \label{fig:sou}}
\end{figure}

\begin{figure}
\includegraphics[width=84mm]{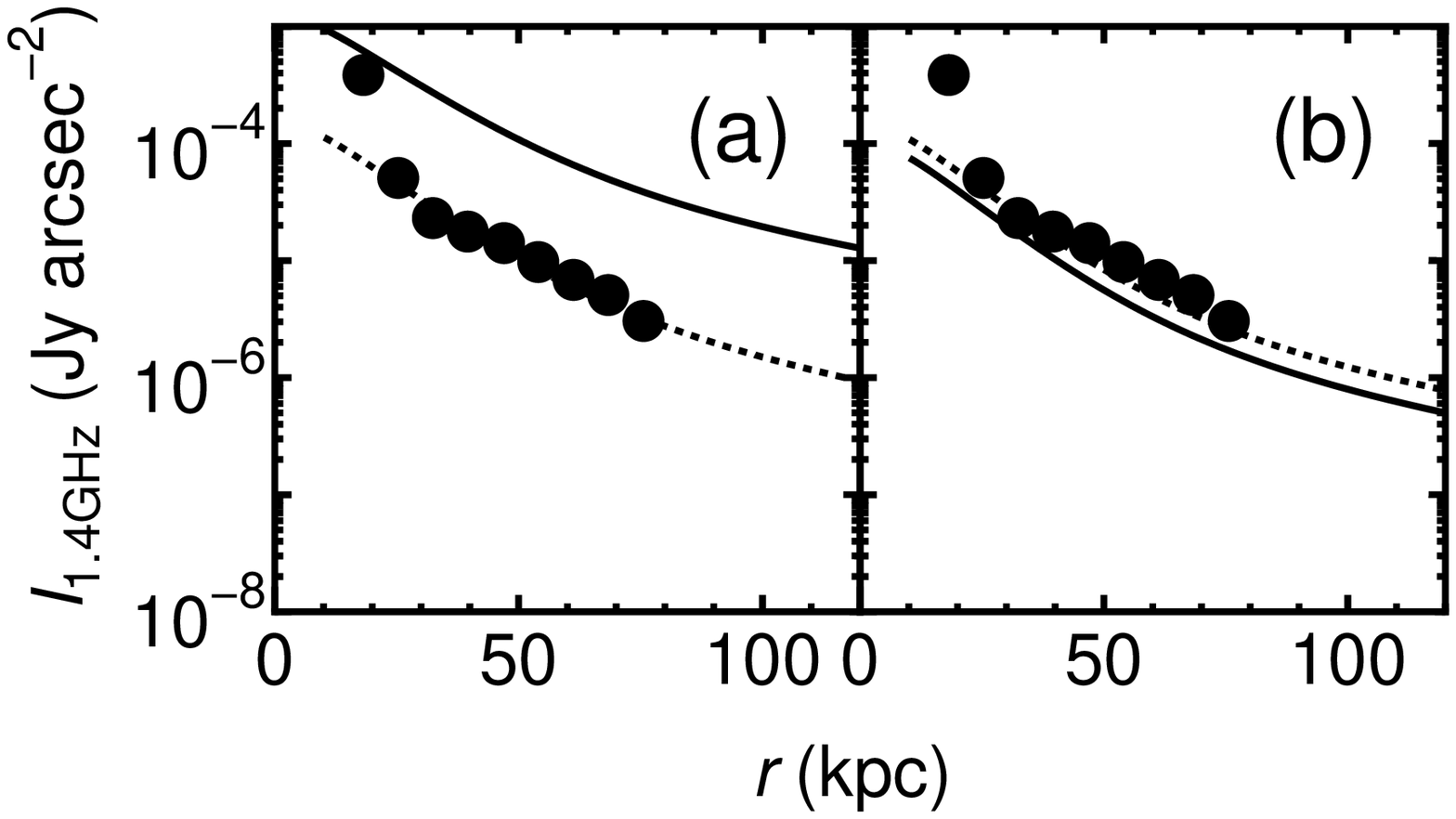} \caption{Profiles of synchrotron
radiation of the Perseus cluster at $\nu=1.4$~GHz. (a) Streaming
velocity is Alfv\'en velocity, $v_A$. The solid line shows $x=2.9$, and
the dotted line shows $x=3.5$. (b) $B=3\: (n_e/0.016\rm\:
cm^{-3})^{2/3}\rm\: \mu G$. The solid line shows $x=2.9$, and the dotted
line shows $x=2.8$. Filled circles are the observations
\citep{ped90a,mur09a} \label{fig:com}}
\end{figure}

\subsection{Structure of the magnetic field}
\label{sec:struc}

In our model, CRs are scattered by Alfv\'en waves. We have implicitly
assumed that the gradient of $P_c$ and the background magnetic field are
parallel. However, magnetic field is often thought to be tangled in the
ICM \citep{sar86a}, which may affect the CR streaming. It has been
indicated that even if the gradient of $P_c$ and the background magnetic
field are not parallel, the plasma is unstable and waves are excited by
CRs \citep{bel05,rak08,byk11,rog12}. Thus, we expect that CRs are
scattered by the waves. However, the effective streaming velocity for a
tangled magnetic field is uncertain, and previous studies have not dealt
with this problem \citep[e.g.][]{loe91,guo08}. Although the detailed
study about it is beyond the scope of our study, we investigate the
effect in a simple way.

We assume that the streaming velocity for randomly tangled magnetic
fields is reduced by a factor of $\alpha$ from the one for the radially
stretching magnetic field ($v_{\rm st}=\alpha c_s$). For a given
magnetic field line, forward or backward Alfv\'en waves can
exist. However, when the waves are excited by CRs, they are driven only
in the direction closer to the CR pressure gradient, or outwards in a
cluster in our case \citep[e.g.][]{lon94}. Owing to the scatter by the
waves, a CR particle is not bound by a magnetic field line. On average,
CRs move outwards along with the waves. If the angle between the
magnetic field line and the direction of the CR pressure gradient is
$\theta$ ($\leq\pi/2$), the effective wave velocity in the direction of
the CR pressure gradient may be smaller than the sound velocity by a
factor of $\cos\theta$, and $\alpha$ may be the average of $\cos\theta$:
\begin{eqnarray}
 \alpha&=&\left(\int_0^{\pi/2}d\theta\sin\theta \int_0^{2\pi}d\phi 
\cos\theta
\right) \nonumber \\
 & & /\left(\int_0^{\pi/2}d\theta\sin\theta \int_0^{2\pi}
d\phi\right) \nonumber \\
& = & 0.5\:.
\end{eqnarray}
Figures~\ref{fig:Tn_lcrh3}--\ref{fig:heat_lcrh3} show the results for
the same parameters of Model~2 but for $\alpha=0.5$. We refer to this
model as Model~3. At $t=12$~Gyr, the mass inflow rate is
$\dot{M}=16\:\rm M_\odot\: yr^{-1}$. The results are not much different
from those of Model~2. However, the smaller streaming velocity makes the
escape of CRs slower and increases $P_c$ (Figures~\ref{fig:Pc_lcrh}
and~\ref{fig:Pc_lcrh3}). It appears that the increase of $P_c$ cancels
out the decrease of $v_{\rm st}$ in equation~(\ref{eq:Hst}). Since the
difference between Models~2 and~3 is not significant, we assume that
$\alpha=1$ in the following sections, unless otherwise mentioned.

\section{Comparison with Observations}
\label{sec:comp}

\subsection{ICM models}
\label{sec:ICM}

We study six clusters (Perseus, A1835, A2029, A2390, RXJ1347, and
Ophiuchus) for which the radial profiles of radio mini-halos have been
studied in detail by \citet[][see also the references in the
paper]{mur09a}. We need to find $n_e(r)$, $T(r)$, and $Z(r)$ for the ICM
of these clusters to obtain $P_c(r)$ using equation~(\ref{eq:Pc}).  In
Table~1, we show the ICM profiles we adopted. Some references give the
functional forms. For other references, we fit the observational data
with appropriate functions.

\subsection{Results}
\label{sec:rad}

Using the ICM models in Section~\ref{sec:ICM}, we can estimate $P_c(r)$
by equation~(\ref{eq:Pc}) for $v_{\rm st}=c_s$. Figure~\ref{fig:pp}
shows the ratio of CR pressure $P_c$ to ICM pressure $P_g$ for the six
clusters. The ICM pressures are derived from the ICM models in
Section~\ref{sec:ICM}. The CR pressures at the cluster centres are less
than 10~\%. In Figure~\ref{fig:sou}, we show the predicted surface
brightness profiles of synchrotron radiation for $P_c$ shown in
Figure~\ref{fig:pp}. For these calculations, we need to specify
$B(r)$. In Figure~\ref{fig:sou}, we assume that $B=10\: (n_e/0.016\rm\:
cm^{-3})^{2/3}\rm\: \mu G$. For comparison, we show the observations
\citep{mur09a,ped90a,bac03a,git07c}. As can be seen, the observations
are consistent with $x\sim 2.9$. The observations of all the six
clusters are reproduced fairly well under the simple assumption of
equation~(\ref{eq:balance}) and almost common index $x\sim 3$. In
particular, the predicted slopes of the radial profiles are consistent
with the observations, and the slopes are less uncertain than $x$. As
was discussed in Section~\ref{sec:HC}, the surface brightness may be
overestimated a factor of few outside the core ($r\ga 200$~kpc). Even
so, the agreement between the predictions and the observations is
good. Among the six clusters, the consistency between the prediction and
the observations of A2390 may be the worst
(Figure~\ref{fig:sou}d). \citet{vik06a} indicated that the central
region of the cluster is strongly disturbed by the central AGN, which
might affect the results.

Figure~\ref{fig:com} is the same as Figure~\ref{fig:sou}a (Perseus) but
for $v_{\rm st}=v_A$ (Figure~\ref{fig:com}a) and for a reduced magnetic
field strength (Figure~\ref{fig:com}b). In general, the Alfv\'en
velocity is smaller than the sound velocity in the ICM. Thus, when
$v_{\rm st}=v_A$, the escape of the CRs from the core delays and $P_c$
becomes larger. Accordingly, for a given spectral shape, the synchrotron
emission becomes larger than that in the case of $v_{\rm st}=c_s$ (the
solid line in Figure~\ref{fig:com}a and that in Figure~\ref{fig:sou}a).
However, if we increase the spectral index to $x=3.5$, the observations
can be reproduced (the dotted line in Figure~\ref{fig:com}a). Although
we do not present the result in the figure, we have also studied the
profile for a reduced streaming velocity ($v_{\rm st}=0.5 c_s$). As is
discussed in Section~\ref{sec:struc}, the reduced velocity increases
$P_c$. However, the effect is not significant, and the index of $x=3.0$
can reproduce the observations. In Figure~\ref{fig:com}b, we present the
results when we reduced the magnetic field to $B=3\: (n_e/0.016\rm\:
cm^{-3})^{2/3}\rm\: \mu G$. The synchrotron emission reduces only a
factor of a few for a spectral index of $x\sim 3$ (the solid line in
Figure~\ref{fig:com}b and that in Figure~\ref{fig:sou}a).  If we
slightly reduce the index to $x=2.8$, the observations are reproduced
again (the dotted line in Figure~\ref{fig:com}b). Thus, the uncertainty
of magnetic fields does not much affect the results.

\begin{figure}
\includegraphics[width=84mm]{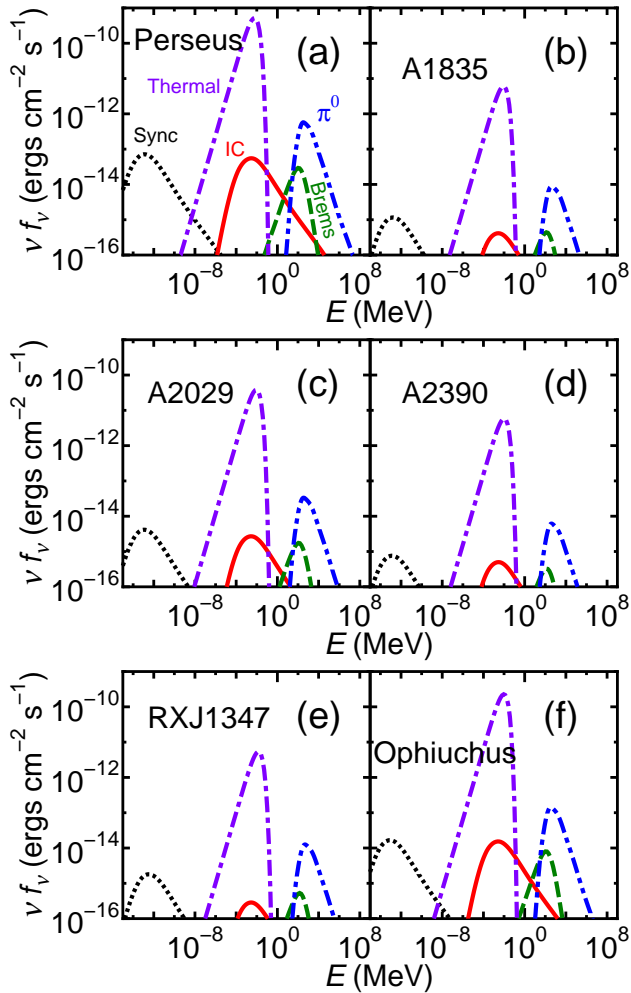} \caption{Broad-band spectra of the
six clusters. Synchrotron radiation (dotted line), inverse Compton
scattering off cosmic microwave background (solid line), and non-thermal
bremsstrahlung (dashed line) are of the secondary electrons. The
$\pi^0$-decay gamma-rays are shown by the two-dot-dashed line. For
comparison, the thermal bremsstrahlung from the ICM is shown by the
dot-dashed line. Redshifts have been corrected.  \label{fig:sp}}
\end{figure}

In Figure~\ref{fig:sp}, we present the broad-band spectra of the six
clusters when $v_{\rm st}=c_s$. We calculated non-thermal emissions from
secondary electrons created through the decay of charged pions that are
generated through the interaction between CR protons and ICM protons. We
also calculated $\pi^0$-decay gamma-ray emissions. We did not include
non-thermal emissions from primary electrons that are directly
accelerated in the ICM, because we have no information on them. We
adopted $x=2.9$ because it can fairly reproduce the radio profiles for
the clusters (Figure~\ref{fig:sou}). Figure~\ref{fig:sp} shows that
except for the Perseus cluster, GeV gamma-ray fluxes are much smaller
than $10^{-12}\rm\: erg\: cm^{-2}\: s^{-1}$, which is the typical
current upper limit set by {\it Fermi} \citep{ack10}. For the Perseus
cluster, the contamination of the central AGN increases the upper-limit
to $\sim 10^{-11}\rm\: erg\: cm^{-2}\: s^{-12}$ \citep{ack10}. Because
of the steep CR momentum spectra, the fluxes in the hard X-ray band and
the TeV gamma-ray band are small. Thus, the detection in these bands
would be very difficult, which is consistent with recent TeV gamma-ray
observations \citep[e.g.][]{ale11}. If non-thermal emissions are
observed in these bands in some clusters in the future, they might come
form primary electrons.

\section{Discussion and Conclusions}
\label{sec:conc}

We have shown that the radial profiles of radio mini-halos observed in
galaxy clusters can be explained by the synchrotron emission from the
CRs that heat the cool cores of the clusters. In our model, the cores
are being heated via CR streaming.

First, we investigated whether the CR heating is balanced with radiative
cooling of the ICM using numerical simulations. We assumed that the
streaming velocity of the CRs ($v_{\rm st}$) is the sound velocity of
the ICM ($c_s$), because it has been pointed out that it is more
appropriate than Alfv\'en velocity ($v_A$) in hot ICM. Moreover, we
found that the heating is more stable when $v_{\rm st}=c_s$ than when
$v_{\rm st}=v_A$. We showed that the CR heating is actually balanced
with the radiative cooling inside the core, although the heating falls
short of the cooling a factor of a few outside the core.

Assuming that the CR heating is balanced with the cooling in the cores
of six real clusters (Perseus, A1835, A2029, A2390, RXJ1347, and
Ophiuchus), we estimated the radial profiles of CR pressure only from
X-ray observations. The results showed that the CR pressure at the
cluster centres are less than 10~\%. Assuming that the momentum spectra
of the CRs are given by a power-law and the CR streaming velocity is the
sound velocity, we calculated radial profiles of the synchrotron
emission from secondary electrons created through proton-proton
interaction. We found that the profiles match the observations if the
index of the CR momentum spectra is $x\sim 3$. If the CR streaming
velocity is the Alfv\'en velocity, the index must be $x\sim 3.5$ to be
consistent with the observations. Uncertainties of magnetic fields in
the ICM do not much affect the results. We have compared our predictions
with the observations of only six clusters, and our model has free
parameters such as $x$. However, if radial profiles of mini-halos are
studied for many other clusters in the future, our model can be examined
more extensively through the comparison with the observations.

We also estimated broad-band spectra for the six clusters when $x\sim
3$. Owing to the steep spectra, hard X-ray and gamma-ray fluxes are
small, which makes it difficult to detect the non-thermal emissions from
the CRs in the clusters in those bands. In the radio band, future
observations of the spectral index will give us information on the
momentum spectra of CRs. For the Perseus cluster, we have shown that our
predicted index is consistent with the observations (Paper~II).

\section*{Acknowledgments}

We thank the anonymous referee, whose comments greatly improved the
clarity of this paper.  We also thank F.~Takahara for useful discussion.
This work was supported by KAKENHI (Y.~F.: 23540308, Y.~O.: 24.8344).

\end{document}